\documentstyle[epsf]{mn} 

\newif\ifAMStwofonts

\ifoldfss

  \ifCUPmtlplainloaded \else
    \NewTextAlphabet{textbfit} {cmbxti10} {}
    \NewTextAlphabet{textbfss} {cmssbx10} {}
    \NewMathAlphabet{mathbfit} {cmbxti10} {} 
    \NewMathAlphabet{mathbfss} {cmssbx10} {} 
  \fi
  \ifAMStwofonts
    \ifCUPmtlplainloaded \else
      \NewSymbolFont{upmath} {eurm10}
      \NewSymbolFont{AMSa} {msam10}
      \NewMathSymbol{\upi}     {0}{upmath}{19}
      \NewMathSymbol{\umu}     {0}{upmath}{16}
      \NewMathSymbol{\upartial}{0}{upmath}{40}
      \NewMathSymbol{\leqslant}{3}{AMSa}{36}
      \NewMathSymbol{\geqslant}{3}{AMSa}{3E}

    \fi
  \fi
\fi 

\ifnfssone
  \newmathalphabet{\mathit}
  \addtoversion{normal}{\mathit}{cmr}{m}{it}
  \addtoversion{bold}{\mathit}{cmr}{bx}{it}
  \newmathalphabet{\mathbfit} 
  \addtoversion{normal}{\mathbfit}{cmr}{bx}{it}
  \addtoversion{bold}{\mathbfit}{cmr}{bx}{it}
  \newmathalphabet{\mathbfss} 
  \addtoversion{normal}{\mathbfss}{cmss}{bx}{n}
  \addtoversion{bold}{\mathbfss}{cmss}{bx}{n}
  \ifAMStwofonts
    \ifCUPmtlplainloaded \else
      \UseAMStwoboldmath
      \makeatletter
      \new@mathgroup\upmath@group
      \define@mathgroup\mv@normal\upmath@group{eur}{m}{n}
      \define@mathgroup\mv@bold\upmath@group{eur}{b}{n}
      \edef\UPM{\hexnumber\upmath@group}
      \new@mathgroup\amsa@group
      \define@mathgroup\mv@normal\amsa@group{msa}{m}{n}
      \define@mathgroup\mv@bold\amsa@group{msa}{m}{n}
      \edef\AMSa{\hexnumber\amsa@group}
      \makeatother
      \mathchardef\upi="0\UPM19
      \mathchardef\umu="0\UPM16
      \mathchardef\upartial="0\UPM40
      \mathchardef\leqslant="3\AMSa36
      \mathchardef\geqslant="3\AMSa3E
    \fi
  \fi
\fi 

\ifnfsstwo
  \DeclareMathAlphabet{\mathbfit}{OT1}{cmr}{bx}{it}
  \SetMathAlphabet\mathbfit{bold}{OT1}{cmr}{bx}{it}
  \DeclareMathAlphabet{\mathbfss}{OT1}{cmss}{bx}{n}
  \SetMathAlphabet\mathbfss{bold}{OT1}{cmss}{bx}{n}
  \ifAMStwofonts
    \ifCUPmtlplainloaded \else
      \DeclareSymbolFont{UPM}{U}{eur}{m}{n}
      \SetSymbolFont{UPM}{bold}{U}{eur}{b}{n}
      \DeclareSymbolFont{AMSa}{U}{msa}{m}{n}
      \DeclareMathSymbol{\upi}{0}{UPM}{"19}
      \DeclareMathSymbol{\umu}{0}{UPM}{"16}
      \DeclareMathSymbol{\upartial}{0}{UPM}{"40}
      \DeclareMathSymbol{\leqslant}{3}{AMSa}{"36}
      \DeclareMathSymbol{\geqslant}{3}{AMSa}{"3E}
    \fi
  \fi
\fi 

\ifCUPmtlplainloaded \else
  \ifAMStwofonts \else 
    \def\upi{\pi}
    \def\umu{\mu}
    \def\upartial{\partial}
  \fi
\fi

\title{Spectral analysis of water vapour in cool stars}
\author[H. R. A. Jones et al.]
       {Hugh R. A. Jones$^1$,
Yakiv Pavlenko$^2$,
Serena Viti$^3$,
Jonathan Tennyson$^3$\\
$^1$Astrophysics Research Institute, Liverpool John Moores University,
Egerton Wharf, Birkenhead CH41 1LD, UK\\
$^2$Main Astronomical Observatory of Ukrainian Academy of Sciences,
Golosiiv woods, 03680 Kyiv-127, Ukraine\\
$^3$Department of Physics and Astronomy, University College London,
Gower Street, London WC1E 6BT, UK\\
}
\date{Accepted  Received 
      in original form }

\date{Accepted ................. Received .............}
\pagerange{\pageref{firstpage}--\pageref{lastpage}}
\pubyear{2000}

\def\LaTeX{L\kern-.36em\raise.3ex\hbox{a}\kern-.15em
    T\kern-.1667em\lower.7ex\hbox{E}\kern-.125emX}

\begin{document}

\maketitle

\label{firstpage}

\begin{abstract}
M star spectra,  at wavelengths beyond 1.35 $\mu$m, 
are dominated by water vapour yet
terrestrial water vapour makes it notoriously difficult to make accurate
measurement from ground-based observations.
We have used the short wavelength spectrometer on the Infrared Space
Observatory at four wavelength
settings to cover the 2.5--3.0 $\mu$m
region for a range of M stars. The observations show a good match
with previous ground-based observations and with 
synthetic spectra based on the Partridge \& Schwenke (1997) line list 
though not with the SCAN (Jorgensen et al. 2001) line list.
We used a least-squared minimisation technique to systematically
find best fit parameters for the sample 
of stars. The  temperatures that we find indicate a relatively hot 
temperature scale for M dwarfs.
We consider that this could be a 
consequence of problems with the Partridge \& Schwenke linelist which leads
to synthetic spectra
predicting water bands which are too strong for a given temperature.
Such problems need to be solved in the next generation 
of water vapour line lists which will extend the calculation
of water vapour to higher energy levels with the good 
convergence necessary for reliable modelling of hot water vapour.
Then water bands can assume their natural role as the primary tool
for the spectroscopic analysis of M stars.

\end{abstract}


\section{Introduction}

More than two-thirds of stars within 10 parsecs are M dwarfs and it is
very probable that this number density prevails throughout our Galaxy.
Unless there is a sharp turn-down in the stellar mass function, they
and even lower mass objects are a major component of the Galaxy's mass.
The dominant red and infrared luminosity of the underlying stellar population
of galaxies is from M giants.
The dominant source of opacity for late-type M dwarfs, giants and brown dwarfs
is water vapour which easily forms in their relatively high pressure,
low temperature atmospheres.  Leaps in theoretical molecular quantum mechanics and computer hardware
capabilities mean that it is possible to perform ab initio calculations to
accurately predict the frequency and intensity for ro-vibrational transitions for
water vapour. This means that it is no longer necessary to extrapolate
laboratory measurements for water vapour to untestable temperature regimes of
which are found in the atmospheres of M dwarfs.

\begin{figure}
\vspace{7cm}
\includegraphics{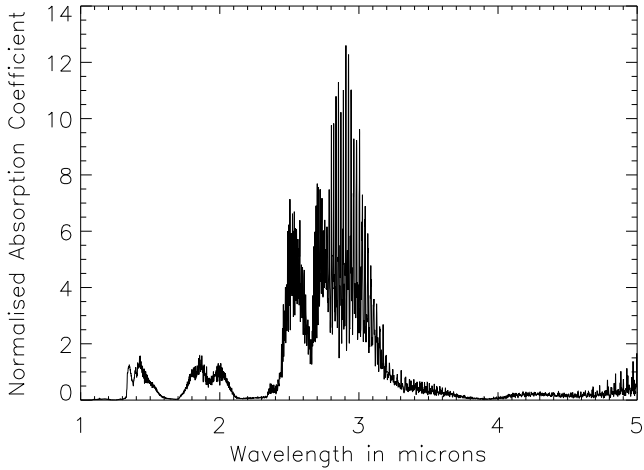}
 \caption{
The water vapour absorption coefficient at 3000 K across the peak of
the flux distribution for late-type M and brown dwarfs. 
The plot has been normalised to one around the maximum absorption of 
water vapour that can reliably be observed from the ground (`high opacity'
case in Jones et al. 1994).
\label{whyiso}}
\end{figure}

The preponderance of water vapour in the Earth's atmosphere makes
it very difficult to observe its spectral signature in stars. At
near-infrared wavelengths, where cool stars emit most of their
flux, the strongest water vapour absorption band is
centred around 2.65 microns where the atmosphere is opaque (Fig.~
\ref{whyiso}). The advent of the Infrared Space Observatory has for the first
time allowed observations to be made at the peak of water vapour
absorption in cool stars. Such data are not only impossible to obtain from
terrestrial sites but also provides a vital overlap with
ground-based data. For M dwarfs this is essential as the data
reduction problems  of decontamination of stellar and terrestrial
water vapour are never far away.

\section[]{Observations}
The spectral region required for this program is inaccessible from
ground-based observatories and is well matched to ISO capabilities. SWS and
ISO provide sufficient resolution to resolve individual ro-vibrational water
bands with enough sensitivity to observe the intrinsically faint and
cool M dwarfs.
The strategy was to observe a sample of M dwarfs which are bright enough
to obtain high signal-to-noise spectra. The ISO programme for these
observations is known as JONES\_PROP32.

We observed a range of M dwarfs - GJ~752B (M8V), GJ~406 (M6V), GJ~699 (M3.5V)
and GJ~191 (M2VI) together with the M4 giant BS~8621. The M dwarfs
were chosen because they have been the subject
of previous studies of M dwarfs (e.g. Jones et al. 1996).
The observations were made in with the Short Wavelength Spectrometer
(SWS) in its full resolution grating mode known as SWS06. Observations
were made using bands 1A and 1B covering the wavelength ranges
2.48--2.60, 2.60--2.75, 2.74--2.90 and 2.88--3.02 $\mu$m. These grating
scans also gave simultaneous coverage from  15.99--16.25 and
16.03--16.21 $\mu$m, however, the relatively low flux levels of the
M dwarfs ($<$10 Jy) and the lower sensitivity of the instrument
means that no useful data were recorded.


\begin{table}
 \centering
 \begin{minipage}{140mm}
  \caption{Properties and total integration times for the sample.}
  \begin{tabular}{@{}llrrrrlrlr@{}}
Object & Flux, Jy (2.76$\mu$m) & Spectral Type &  Time, s \\
      GJ~191  & 4.13$\pm$0.09  & M2.5VI & 4530 \\
      GJ~699  & 6.66$\pm$0.14 & M4V & 1192 \\
      GJ~406  & 1.503$\pm$0.031 & M6V & 9839 \\
      GJ~752B  & 0.019$\pm$0.020 & M8V & 18216 \\
      BS~8621  & 509.7$\pm$4.1  & M4III & 1238 \\
\end{tabular}
\end{minipage}
\end{table}


\begin{figure}
\vspace{7cm}
\includegraphics{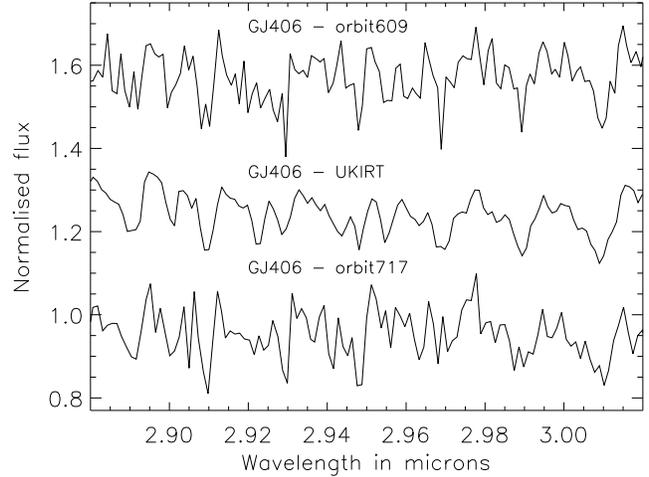}
\caption{A comparison of datasets for GJ~406 from orbits 609 and 717
of this programme and from UKIRT data from Jones et al. (1995).
This object has the lowest  
signal-to-noise ratio in our analysis yet shows a reasonable agreement
between spectra taken on different orbits and from the ground. 
\label{comp}}
\end{figure}

\begin{figure*}
\vspace{8cm}
\includegraphics{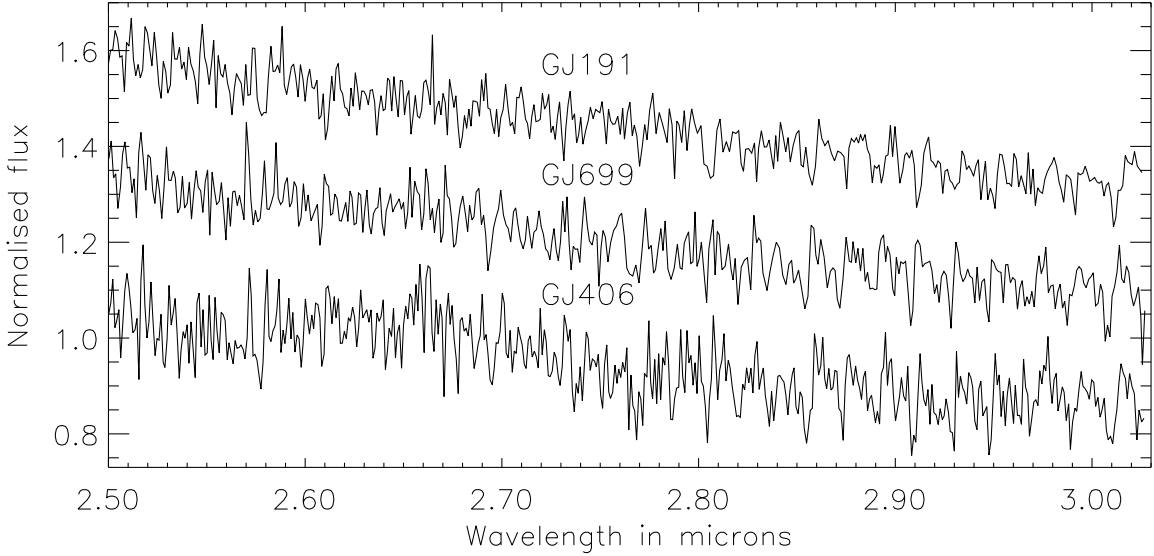}
\caption{A spectral sequence for GJ~191 (M2VI), GJ~699 (M4V) and GJ~406 (M6V).
\label{SPU_fig2}}
\end{figure*}

\begin{figure*}
\vspace{8cm}
\includegraphics{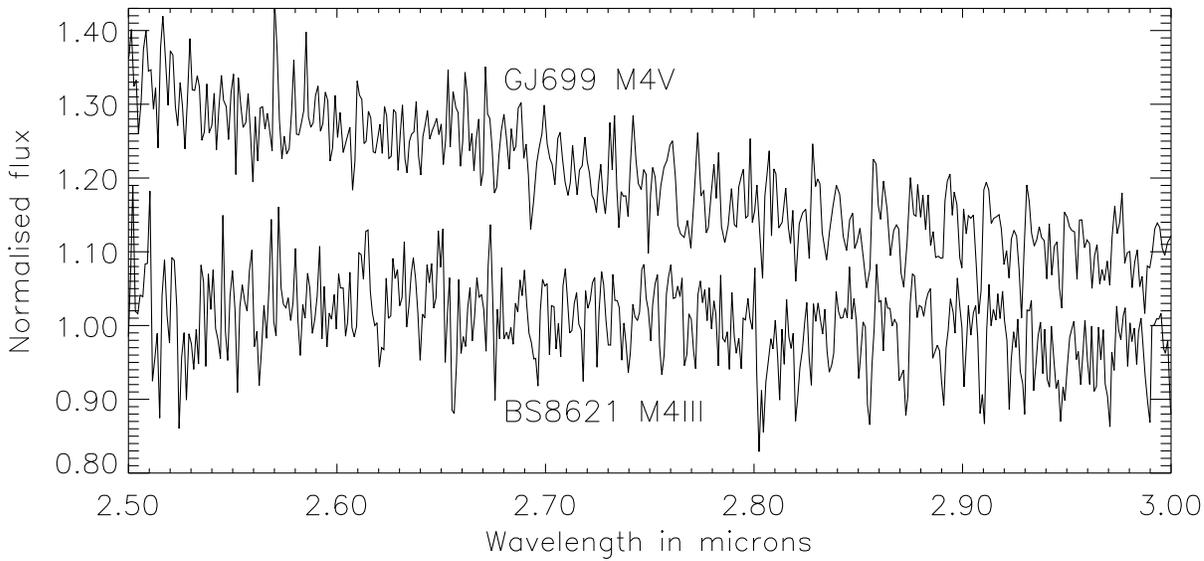}
\caption{
A comparison at optical spectral type M4 for the dwarf (V)
GJ~699 and the giant (III) BS~8621.
\label{SPU_fig3}}
\end{figure*}


\section{Data reduction}

The observational data were processed within the Observers SWS
Interactive Analysis Package from the standard processed
data to Auto Analysis Result level through the
standard pipeline of Derive Auto Analysis Product. In addition to
a few bad data points flagged by the auto analysis software ISO,
SWS data also contains glitches and jumps not easily corrected for
by the standard reduction tools (Heras 1997). 
The philosophy adopted for the data
presented here was to ignore all scans effected by glitches or
jumps and to let individual bad points be taken care of by sigma
clipping. The glitches and jumps do not tend to appear in the
brighter objects though for the fainter objects, e.g., GJ~406, they
affect around 10 percent of scans. The proc\_band and div
procedures were used to print out scans for each individual
detector for each observation for all targets. These were visually
inspected to check for glitches. and bad data points. The ISO
Spectral Analysis Package (ISAP) was used to remove bad data
points, check the data and to produce the final combined spectra
by averaging across all detectors using the standard clip mean
option. We found little sensitivity to the method used to combine
the data.

We also made observations of the archetypal late-type M dwarf GJ~752B.
These data are tantalising however our conservative
reduction philosophy means that little of this data survives.
It would be possible to flag all glitches and jumps, however, they
are usually accompanied by gradients affecting most of the data in
a scan.
We look forward to the advent of reduction tools that will be
able to deal with flux levels at the 0.01 Jy level.

The calibration files used for the reduction were those 
included in version 8.4 of ISAP package. Based
on comparisons between objects taken with the same configuration
during different orbits the flux calibration varies by only a few
percent. This is expected for band 1 SWS observations which have
excellent calibration. The wavelength calibration for the SWS
instrument is measured to be within 1/8 of a resolution element
(Salama et al. 2000). The available ground-based data for our
sample are at lower resolution than the ISO data, nonetheless,
bear out the calibration of the SWS data. Fig.~\ref{comp} shows the
good agreement in the region of overlap between parts of the
dataset taken on different orbits and ground-based measurements
(from Jones et al. 1995).

\begin{figure}
\epsfxsize=8cm \epsfysize=8cm \epsfbox{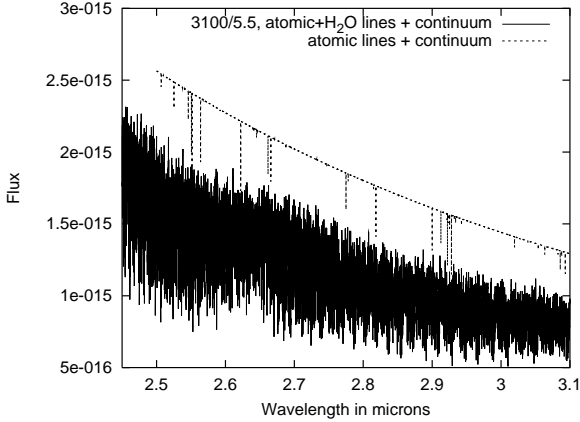}
\caption{Theoretical SEDs computed for 3100/5.5 model atmosphere
taking into account continuum + atomic line and continuum +
atomic lines + H$_2$O opacities \label{3155a-m}}
\end{figure}

\section{Available water linelists}

The crucial opacity for M stars across the range of our observed spectra
is water vapour (e.g., Fig.~\ref{3155a-m}). A useful model atmosphere 
thus needs to include an appropriate water linelist. There are 
several different sources of water data and a few of them 
have been investigated here.  
A reliable water linelist needs: 1. a good (electronic) potential 
energy surface; 2. well converged nuclear motions (i.e vibration-rotation) 
calculations; 3. a reliable dipole surface. To date, it is not yet possible
to a get a completely reliable {\it ab initio} potential surface, so 
all the linelists discussed in this paper used surfaces which have been
adjusted to reproduce laboratory spectroscopic data for water.
Fitting to lab data can cause problems in regimes where such 
data are unavailable (Polyansky et al. 1997a). 
Conversely tests (Lynas-Gray, Miller \& Tennyson 1995) 
have shown that {\it ab initio}
dipole surfaces are much more reliable than the ones fitted to experimental
data.
\par
The water vapour data investigated in this paper
are from
Miller et al. (1995, known as  the MT list), from
Partridge \& Schwenke (1997, known as the PS or AMES list), from 
Jorgensen et al. 2001 (known as SCAN), and from
Viti (1997, known as the VT2 linelist) which superseded
the MT and VTP1 line lists (Viti, Tennyson \& Polyansky 1997). 
MT is a relatively small list (10 million lines) which used,
by today's standards, a rather inaccurate potential. VTP1 energy surface is
much more accurate than MT but it is still not complete enough for
stellar models. VT2 is a much larger linelist ( $>$ 300 million lines) 
which uses
a potential surface reliable for higher vibrational states. However, due to
serious computer format corruption problems this list could not be 
included in model atmospheres calculations.

\begin{figure}
\vspace{7cm}
\includegraphics{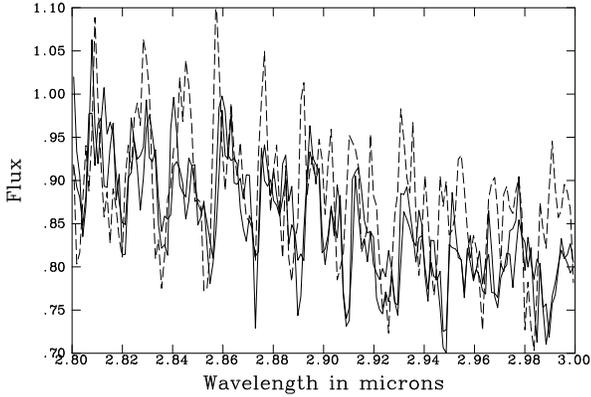}
\caption{GJ~406 compared with 3000 K synthetic spectra using the MT
and PS linelists. The observed spectra and PS model are shown as
solid thick and thin lines respectively and the MT model as a dotted
line. The match of water bands between observations and model is
considerably better for the PS model.}
\label{mtps}
\end{figure}

\begin{figure}
\vspace{7cm}
\epsfxsize=8cm \epsfysize=8cm \epsfbox{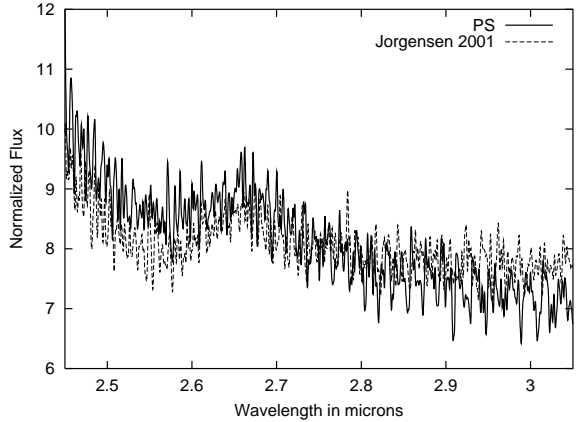}
\caption{Comparison between synthetic spectra at 3100 K using the SCAN
and PS line lists.}
\label{psjorg}
\end{figure}

\section{Spectral analysis}

For this analysis we computed spectral energy
distributions for a range of model atmosphere parameters:
effective temperatures, $T_{\rm eff}$ = 2700 --- 3900 K, metallicities
[Fe/H] = 0.0 --- --1.5 and gravities log $g$ = 0.0 --- 5.5.
From here on we refer to these as
model synthetic spectra.
These were generated using the temperature structures
from the model atmospheres of Hauschildt, Allard \& Baron (1999)
by the WITA6 program (Pavlenko 2000).
In our computations we used the atomic line list from VALD 
(Kupka 1999), the
water vapour line lists of Partridge \& Schwenke (1998)
and Jorgensen (2001) and the CO line list of 
Goorvitch (1994).

We also used synthetic spectra from the PHOENIX
model atmosphere code, the so called NextGen version 5
(Hauschildt et al. 1999).

We adopt the Voigt funtion $H(a,v)$ for the line shape of any line 
in our computations. 
Damping constants
$a = (\gamma_2 + \gamma_4 + \gamma_6)/(4\times\pi\times\Delta\nu_D)$ 
are determined in atmospheres of cool dwarfs mainly by 
Van der Waals (pressure) broadening ($n$=6) and (in upper layers)
natural broadening ($n$=2). Stark 
broadening ($n$=4) of spectral lines can be 
neglected here due to low temperatures and electron 
densities. For atomic lines the line broadening constants were taken 
from VALD. Unfortunately, for molecular line broadening models 
are rather uncertain. We assume that the 
mechanisms of broadening molecular and atomic lines are 
similar.   
Radiative broadening of molecular lines were computed in the 
frame of classical approach $\gamma_2 = 
0.22/\lambda^2$ (Allen 1973). Due to the high pressures and low 
temperatures in the atmospheres of cool dwarfs Van der Waals 
broadening should dominate there (see Pavlenko 2001).
The damping constants of pressure broadening of molecular 
lines $\gamma_6$ were computed following Unsold (1948).  

To provide an idea of the improvement since our previous work 
(Jones et al. 1995), in Fig. ~\ref{mtps} we compare GJ~406 with 
3000 K synthetic spectra generated using the PS and MT lists. As 
with Jones et al. (1995) we found that the intensities of the 
water band strengths seem to be reasonably well determined. 
However, we find that the PS line list prediction of bands with 
wavelength is a substantial improvement on the previously used MT 
line list. We also investigated using the SCAN line list 
(Jorgensen et al. 2001) and in Fig. ~\ref{psjorg} compare this to 
the PS list. As with our previous comparisons to the SCAN list 
(Jones et al. 1996) we find the predicted shape of the water 
bands as well as the structure of the bands are rather different 
from other models and observations. As expected from the scale of 
the calculation and previous comparisons with observations (e.g., 
Allard et al. 2000) we find that the PS line list is the most 
useful water line list available for our observed data. We first 
proceed to investigate the completeness of the PS line list and 
then use it to find best-fit parameters for our targets. 

\begin{figure}
\vspace{6cm}
\includegraphics{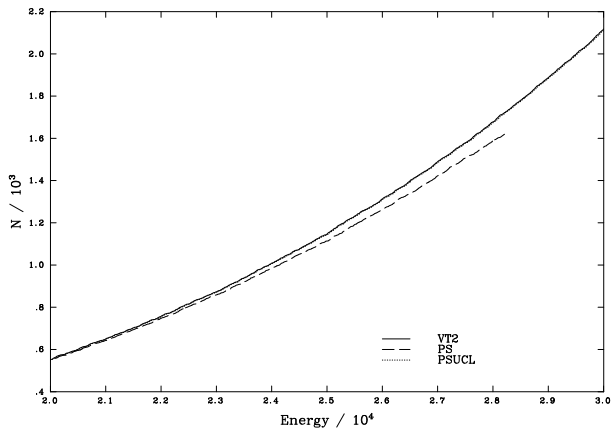}
\caption
{Number of energy levels as a
function of energy for PS (dotted) and VT2 (continuous) linelist for
$J$ = 17. The PS linelist stops at 28000 cm$^{-1}$.}
\label{psvs17}
\end{figure}

\subsection{Completeness of the PS line list?}

For model atmospheres with effective temperatures in the range
2000 --- 4000 K, Boltzmann considerations suggest that transitions
involving states with J = 20 --- 30 are the dominant sources of
opacity.
We find two problems with the
PS linelist: (i) first of all, it does not reach 30000 cm$^{-1}$ for
any $J$ apart from $J$ = 0. On average it gets to 28000
cm$^{-1}$. (ii) Secondly, even by truncating energy levels to 28000
cm$^{-1}$, VT2 still has more levels than PS has for high $J$, for
example, at $J$ = 17, VT2 gives 100 energy levels more than PS below
28000 cm$^{-1}$.  From the number of energy levels missing for
each $J$ we estimate that PS omits $\sim$ 30 $\%$ transitions up
to $J$ = 10; $\sim$ 50$\%$ up to $J$ = 20; and $\sim$ 60$\%$ up to $J$
= 28.

As a test, 
a new set of energy levels
using the PS potential energy surface, the same atomic masses as
PS  and VT2 nuclear motion parameters, were computed.
This new set of energy levels is known as UCLPS
(University College London Partridge Schwenke list; 
see Polyansky et al. 1997b for further comparisons).
This was done to
test the variational convergence for the same potential energy
surface.  An example of the energy levels comparison with VT2 and the new
calculations is shown for $J$ = 17 in Fig.~\ref{psvs17}.
The calculations performed
with the PS potential energy surface are in close agreement with VT2
which suggests that the lack of states in the PS linelist is due to
poor convergence and not to differences in the potential energy
surface used.  In fact, close examination of the parameters of the PS
calculation strongly suggests that their decision not to increase the
size of the Hamiltonian matrix beyond that used for $J$ = 4 resulted in
poorly converged calculations for higher states with high $J$. This
decision undoubtedly saved them from some of the computational
problems experienced in the computation of the VT2 linelist
(Viti 1997).

For low-J energy levels,
PS's calculations give superb results, reproducing experiment
with a much higher accuracy than than VTP1 or VT2. However, for
higher rotational states, particularly those with $J >$ 20,
we find that a very high proportion of rotational states which
ones expects to be degenerate in fact show significant splittings
in the PS linelist. This splitting is not shown in VTP1, VT2 or
in UCLPS. In particular for the levels with high $K_a$
and
with $K_c$ odd (where $K_a$ and $K_c$ are the projections of $J$ on the 
A and C principal
axes of rotation of asymmetric top molecules),
all lie below the ones with $K_c$ even for levels
with which they should be quasi-degenerate. Since the PS
calculation truncates variational rotation-vibration calculations
with 7500 energy-selected basis functions independent of the
rotational parity this means that the $K_c$ odd
calculations will contain states of higher cut-off energy than the $K_c$
even
calculation. The variational principle means
that the $K_c$ odd states will be better converged and hence lower
in energy. 

This causes two problems. Firstly,
artificial splitting of lines means that it is
difficult to use the list for line assignments (Polyansky et al. 1997a).
The second is more subtle.
An important consideration in radiative transport is
how the line absorptions fill in gaps in the spectrum.
Two transitions which, to within their linewidth, are coincident
will have less effect on the opacity than two
separate transitions.  Artificially
doubling the number of lines for these J values is likely
to cause the strength of water vapour bands at low resolution
to be overestimated.

It should be noted that the resolution
of the observed spectra (see Fig.~\ref{comp})
is much too low to see the rotational structure of H$_2$O
shown in the theoretical spectrum (Fig.~\ref{3155a-m}).
Although the calculated line positions of the PS list agree well
enough with the experimental ones (tabulated, for example,
in HITRAN96, Rothman et al. 1996), individual line strengths
can differ greatly. 
Recently Schwenke \& Partridge (2000) found problems with their
analytical representation of the {\it ab initio} dipole moment data used
for the computation of the PS linelist: this may have led to an
overestimation of the intensities of weak bands with respect to experimental
data. They improve their previous analytical representation by careful and accurate
fitting. However their new dipole surface has
yet to be used to produce an improved water linelist.  

\begin{figure}
\epsfxsize=8cm \epsfysize=8cm \epsfbox{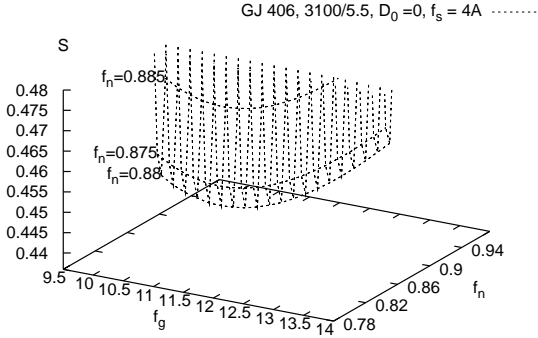}
\caption{An example of the dependence of $S$ on $f_{\rm n}$ and $f_{\rm g}$, 
values of $f_{\rm n}$ and $f_{\rm g}$ are in ${\rm\AA}$. 
\label{sens-fn}}
\end{figure}

\begin{figure}
\epsfxsize=8cm \epsfysize=8cm  \epsfbox{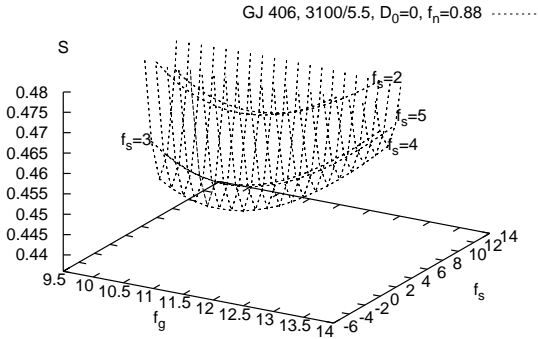} 
\caption{An example of the dependence of $S$ on $f_{\rm s}$ and $f_{\rm g}$,
values of $f_{\rm s}$ and $f_{\rm g}$ are in ${\rm\AA}$. 
\label{sens-fg}}
\end{figure}

\subsection{Best-fit parameters}

Our synthetic spectrum models
were calculated at high resolution, typically 0.5 --- 1
${\rm\AA}$. Before comparison with model spectra were made
the observed data were corrected for their radial velocities:
244.0, 110.9 and 21.0 km/s respectively for GJ~191, GJ~699 and
GJ~406 (Leggett 1992). The model were transformed to the resolution 
of the data and to wavelengths in air. The resolution of ISO data 
are actually somewhat uncertain
and have been the subject of a number of studies (e.g. Valentijn et al. 1996,
van den Ancker et al. 1997, Lutz,
Feuchtgruber \& Morfill, 2000). The SWS06 mode is a grating scan
mode and in principle reproduces the full intrinsic
resolution of the grating (so-called SWS02 mode).
However, the SWS06 mode has poor sampling of 
a given wavelength
element. For the purpose of the derivation of high-fidelity line
profiles, this causes unwanted interactions between line profiles
and the relative sensitivities for the 12 detectors of a band
('flat-fielding in SWS-speak'), since all twelve detectors have to
be combined for a good profile. This means that although the
measured resolution of SWS02 in band 1A (2.38 --- 2.60 $\mu$m is
0.0007$\mu$m and 0.0011$\mu$m in band 1B (Lutz et al. 2000), the
actual resolution of these SWS06 spectra is likely to be somewhat less.

In order to determine the best fit parameters for our targets in 
a systematic fashion, we performed a least-squared minimisation
on the spectra within a grid of synthetic spectra using the PS line list.
As with the previous
work we noticed that the form of the observed spectra along 
with its unresolved nature means that the line broadening 
employed by the models has a substantial influence on the fits.
Although the SWS line profile is well determined to within 1 per
cent, the resolution of the data are not so secure (Lutz et al. 2000).
To find the best fit model parameters we allowed the spectral 
normalisation, wavelength shift and resolution to vary. 

We compared observed fluxes $H^{\rm obs}_{\rm \lambda}$ with
computed values $f_{\rm n} * H^{\rm theor}_{\rm \lambda + f_{\rm s}}$.
We let $H^{\rm obs}_{\rm lambda}$ = $\int (F^{\rm theor}_{x-y}*G(y)*dy$,
where the theoretical flux is $F^{\rm theor}_{\rm \lambda}$ and
$G(y)$ is the instrumental profile modelled by a Gaussian. 
In our case $G(y)$ may be wavelength dependent.
To get the best fit we find the minima per point of
the 3D function

$S(f_{\rm n},f_{\rm s},f_{\rm g})=\Sigma(1-H^{\rm synt}/H^{\rm obs})^2$.

We calculated this minimisation with each observed spectrum 
and the grid of synthetic spectra to determine a set of parameters 
$f_{\rm n}$ (normalisation factor), $f_{\rm s}$ (wavelength shift parameter), 
$f_{\rm g}$ (resolution) for all the stars of our sample.  

In general we found the
dependence of $S$ on  $f_{\rm n}$ and $f_{\rm s}$ is much stronger than on
the $f_{\rm g}$. For example see Fig.~\ref{sens-fn} and \ref{sens-fg} 
for the case GJ~406. It should be noted that although the resolution
across a grating setting is expected to be the same the resolution
is expected to decrease slightly between the so called band 1A 
(2.48 --- 2.60 $\mu$m) and band 1B (2.60 --- 3.02 $\mu$m) data. We account
for this by also investigating the band 1A and band 1B data separately.
Best fit values for $f_{\rm g}$ were investigated for the observed spectra,
across most of the grid of model atmospheres.  They  
ranged from 7.0 (band 1A) to 18.2 (band 1B) ${\rm\AA}$ and always showed 
the expected resolution
difference between band 1A and 1B data, e.g.,  example minimisation
plots are shown for GJ~406 and a 3100 K, log $g$ = 5.5, solar metallicity
model in  Figs~\ref{sens-fn} and \ref{sens-fg}. 
We generally found the largest and smallest values for $f_{\rm g}$ when
making comparisons with synthetic spectra with properties far from
the best fit minimisation $S$ values. For example in the case of 
GJ~406, the extreme best fit $f_{\rm g}$ values of 8.0 and 18.2  ${\rm\AA}$  are 
obtained with synthetic spectrum models of 3300 K, [Fe/H]=--1.5 and 2700 K, 
[Fe/H]=0.0 respectively. Apart from giving us an independent check of
the resolution, this procedure served to investigate the sensitivity of
minimisation $S$ values to resolution $f_{\rm g}$. Although different
$f_{\rm g}$'s give significantly different values of $S$ we found that
the best fit temperature for a given $f_{\rm g}$ was stable.
We also found $S$ values for $observed$ and synthetic spectra 
smoothed by the same effective values. Although this gave the same best fit 
minimisation synthetic spectra, this conservative solution involves 
smoothing the observed data and led to a decrease 
in the sensitivity of the best fit solution.
Given the numerical stability of our best fits to resolution and the lack
of evidence for any changes in the SWS resolution with time we felt
confident using the adopting the resolution values of 0.0012 and 0.0017 $\mu$m
for band 1A and band 1B data as given in the SWS Observer's 
manual (de Graauw et
al. 1996).

Our minimisation procedure also yields normalisation values
$f_{\rm n}$ for the different spectral regions. 
We find that the band 1A data are fainter than the
band 1B data by 1--2\% using our best fit synthetic 
spectra. Given that this is generally better than the formal signal-to-noise
of our spectra and that this result is somewhat model dependent we thus 
consider the flux calibration of the data to be robust and do
not apply any new flux calibration to the data.
The minimisation procedure also yields values for the
wavelength shift $f_{\rm s}$. We found a shifts of 
3.7--4.3 ${\rm\AA}$  between observed and synthetic spectra which 
corresponds to around 0.3 of our data resolution. This shift
maybe due to the wavelength calibration of the data though at
least part may arise from inaccuracies in the
water vapour line list used to generate the synthetic spectra. 

Our results are also sensitive to the rather uncertain damping parameters
adopted for molecular lines. To test this dependence two models at
2700 K, log $g$ = 5.5, [Fe/H]=--0.5 with different damping constants 
were compared. The literature for atomic lines (e.g., Gurtovenko \& 
Kostyk 1989) tells us that that 
a factor of two might be the maximum expected for this uncertainty due
to the difference
between classical and quantum mechanical computations. Although 
both atomic and molecular lines are damped osciallations the 
oscillations are greater in molecules so a larger factor maybe appropriate. 
Nonetheless it is also possible that molecular lines are less sensitive
to damping because the structure of molecular
levels maybe more flexible to changes of the surrounding
electric fields: electrons of molecular orbitals are less bound
in comparison with atoms. There may be something like
a ``soft (quasi-)adiabatic response''.
We conducted a rather conservative experiment and compared models using 
Unsold's $\gamma_6$ and with 5$\times\gamma_6$ 
and found a flux difference of around 5\%. 
For our comparisons we consider this is an
upper limit as in all cases we are investigating weaker molecular
bands, due to higher effective temperatures and/or lower metallities.
Furthermore, we are investigating the spectral differences
in terms of synthetic spectra at relatively low resolution 
and find the expected uncertainty due to molecular 
damping constants to be much less than 1\% and so not
significant for our results.

The following subsections investigate our fits to the
each observed spectra in turn.
The best fit models are then compared with a compilation of values
from the literature in Section~5.6. 

\begin{figure}
\epsfxsize=8cm \epsfysize=7cm \epsfbox{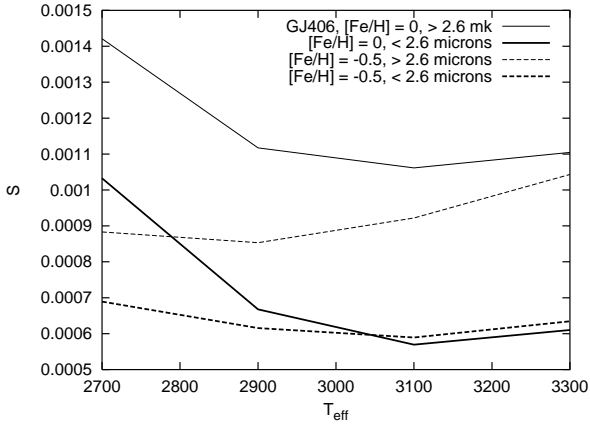}
\caption{
Minimisation $S$ for different synthetic spectra compared to GJ~406
spectrum. 
\label{cases-fg}}
\end{figure}

\begin{figure}
\epsfxsize=8cm \epsfysize=8cm \epsfbox{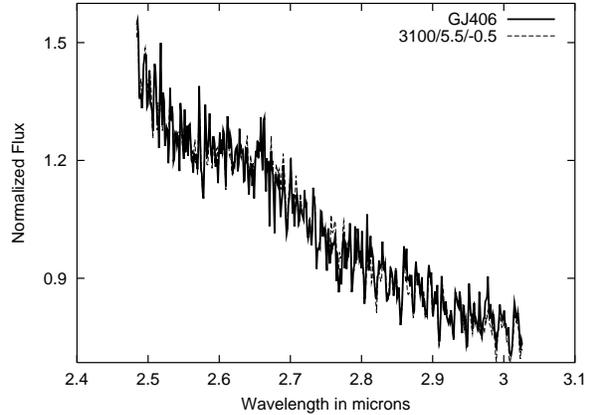} \caption{
The observed spectrum of GJ~406 compared to a 3000 K, log $g$ = 5.5,
[Fe/H]=--0.5 synthetic spectrum.
\label{fit406}}
\end{figure}

\subsection{GJ 406}

As discussed above and as illustrated in Figs~\ref{sens-fn} and 
\ref{sens-fg} we carried a number of different tests on GJ~406 
in order to find our preferred model fits. We tended to use 
GJ~406 as our primary observed spectrum because of its late spectral
type and strong water bands. In addition to the tests discussed above
GJ~406 has data taken during four different orbits which allows 
us to analyse the orbits separately, thus checking any time 
dependence problems with the dataset. We found differences in the 
minimisation values obtained of around 7\% for the 2.48---2.90 $\mu$m
data and around 20\% for the 2.88---3.02 $\mu$m data though the best
fit models were all within 100 K.     

Fig.~\ref{cases-fg} illustrates the minimisation, $S$ for various
different synthetic spectra with the observed GJ~406 spectrum.
For this plot we have split up the fits into those for 
data shortward and longward of 2.6 $\mu$m. The improved
minimisation for the higher resolution $<$2.6$\mu$m data can be
seen. As can the rather complex sensitivity of the minimisation 
with temperature and metallicity. The best fit value appears 
to be for [Fe/H] = --0.5 and 3000 K with the solar metallicity
models suggesting 3100 K and [Fe/H] = --1.0 models suggesting 2900 K.
An illustration of the quality of the fit can be seen 
in Fig.~\ref{fit406}

\begin{figure}
\epsfxsize=8cm \epsfysize=8cm \epsfbox{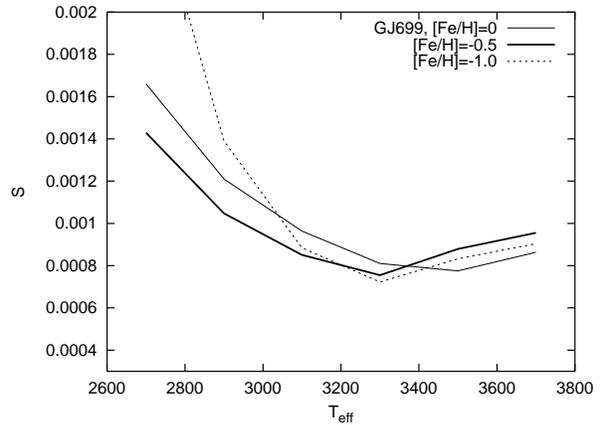} \caption{
Minimisation $S$ for different synthetic spectra compared to GJ~699
spectrum. 
\label{s699}}
\end{figure}

\begin{figure}
\epsfxsize=8cm \epsfysize=8cm \epsfbox{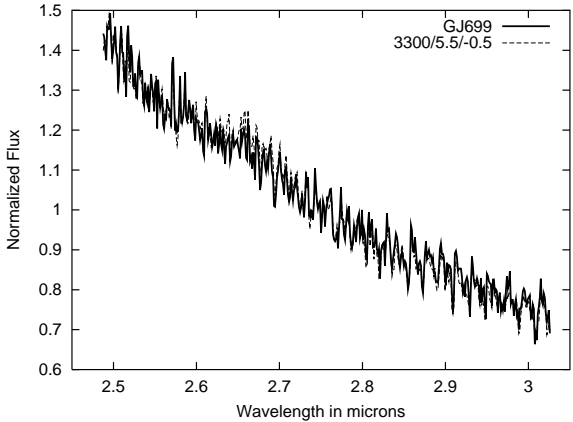} \caption{
The observed spectrum of GJ~699 compared to a 3300 K, log $g$ = 5.5,
[Fe/H]=--0.5 synthetic spectrum.
\label{f699}}
\end{figure}

\subsection{GJ 699}

GJ~699 (Barnard's star) is the largest proper motion object known outside
our solar system and is thus suspected of being relatively old and having
a low metallicity. In fact it tends to skew the proper motion properties
and characteristics of the local neighbourhood so much that it is 
sometimes ignored
from such studies. Nonetheless it is the most nearby and  
brightest known M4V dwarf and thus was included as target for this
programme. Fig.~\ref{s699} shows the minimisation plot for the observed
spectrum.
This suggests a best fit synthetic spectrum with 3300 K and [Fe/H]=--0.5.
In Fig.~\ref{f699} we compare this fit with the observed spectrum.

\begin{figure}
\epsfxsize=8cm \epsfysize=8cm \epsfbox{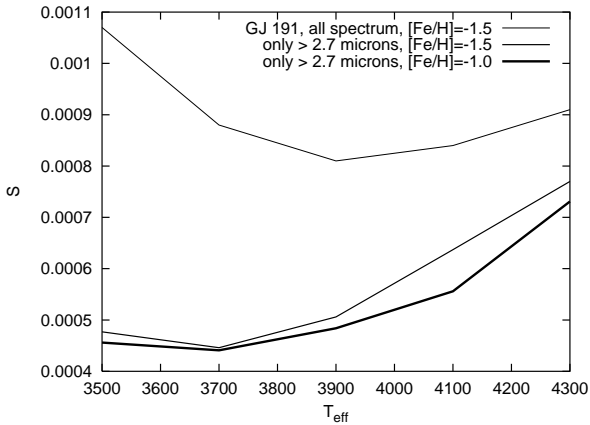} \caption{
Minimisation $S$ for different synthetic spectra compared to GJ~191
spectrum.}
\label{s191}
\end{figure}

\begin{figure}
\epsfxsize=8cm \epsfysize=8cm \epsfbox{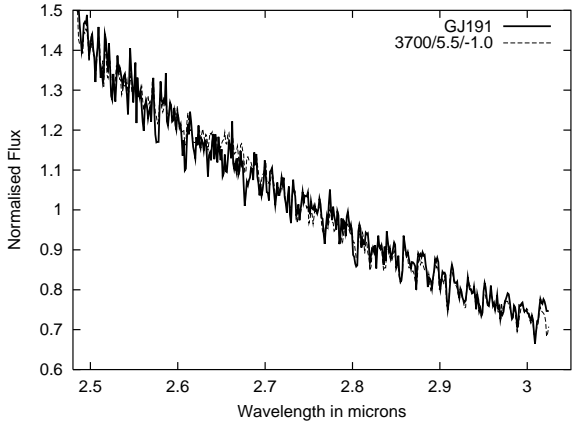} \caption{
The observed spectrum of GJ~191 compared to a 3700 K, log $g$ = 5.5,  
[Fe/H]=--1.0 synthetic spectrum.
\label{f191}}
\end{figure}

\subsection{GJ~191}

GJ~191 has a high space velocity and displays CaH bands
in its optical spectra and unlike GJ~699 is indisputably 
spectrally typed as a subdwarf. 
Our analysis of $S=f({\rm [Fe/H]},{\rm T_{eff}})$, 
e.g., Fig.~\ref{s191}, showed a best
fit temperature of 3700 K and  [Fe/H] 
values in the range --1.0 --- --1.5.
Computations with larger/smaller  [Fe/H] provide
{\em systematically} larger $S$. 
Unlike GJ~406 and GJ~699 we found that the match
of observed and synthetic spectra was considerably better
for data longward of 2.7 $\mu$m. We suspected that this
might be caused by the interference of CO $\Delta$$\nu$~=~2
bands, so we investigated higher (log $N$(C)~=~--3.13) and lower
(log $N$(C)~=~--3.68) carbon adundances than the scaled solar value 
(log $N$(C)~=~--3.48), however, we found no evidence that 
the discrepancies are caused by CO. Since the 
incompleteness of the water line list increases with 
increasing temperature (e.g., Fig.~\ref{psvs17}) 
it is to be expected that the
spectrum of GJ~191 is less well fit than the cooler dwarfs.
Allard et al. (2000) reach a similar conclusion using the 
PS line list: `The introduction of the PS-H2O opacities brings 
solid improvements of the near-infrared SED of late-type dwarfs 
but fails as the AH95 models did to reproduce adequately the J-K 
colors of hotter stars'. Despite these problems a reasonable match 
between the 
observed and synthetic spectra can be seen in Fig.~\ref{f191}.

\begin{figure}
\epsfxsize=8cm \epsfysize=8cm \epsfbox{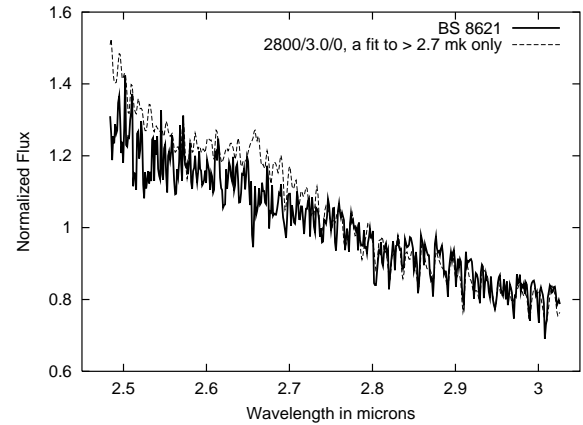} 
\caption{The
observed spectrum of BS 8621 compared to a 2800 K, log $g$ = 3.0,
solar metallicity synthetic spectrum.
\label{f8621}}
\end{figure}

\begin{figure}
\epsfxsize=8cm \epsfysize=8cm \epsfbox{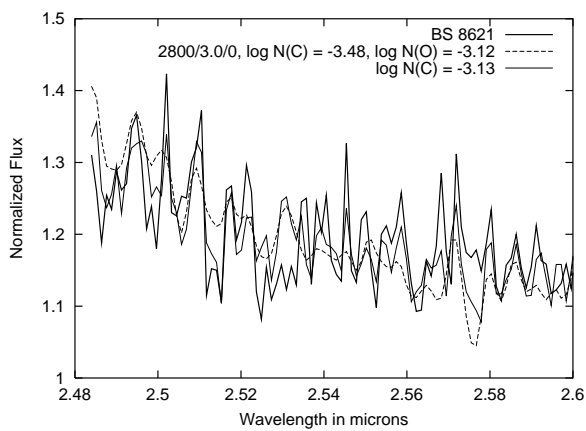}
\caption{The
observed spectrum of BS 8621 compared to 2800 K, log $g$ = 3.0
synthetic spectra with two different carbon
abundances. The influence of an enhanced carbon abundance can 
be seen prominently on the CO bands between 2.49 and 2.52 $\mu$m.
\label{ff8621}}
\end{figure}

\begin{table*}
 \centering
 \begin{minipage}{140mm}
  \caption{Best fit models for the sample and a collection of
effective temperatures and metallicities from the literature.
The value for BS~8621 is given without an 
error to reflect the uncertainty in this value.
References are as follows
(1) Krawchuk, Dawson \& De Robertis (2000), (2) Gizis (1997),
(3) Leggett (1992), (4) Cayrel de Strobel et al. (1985), (5) Bessell (1991),
(6) Tinney et al. (1993), (7) Jones et al. (1994), (8) Leggett et al. (2000,
apj, 535, 965), (9) Reid \& Gilmore (1984), (10) Veeder (1977)
, (11) Pettersen (1980), (12) Mould (1976), (13) Basri et al (2000), (14) Kirkpatrick
et al. (1993), (15) Tsuji et al. (1996), (16) Berriman et al. (1992), (17)
Brett (1995), (18) van Belle et al. (2000).
}
  \begin{tabular}{@{}lllll@{}}
Object & Best fit model &  Values of $T_{\rm eff}$ (and [Fe/H]) from
the literature (Reference given in caption)\\\\

      GJ~191  & 3700$\pm$100 K,&
3466 --1.82(1), 3700 --1.5(2), -0.5 (4), 3550 -0.55 (12)  \\
& [Fe/H]=--1.0$\pm$0.3 & \\
      GJ~699  & 3300$\pm$100 K,&
2924 --1.35 (1), 3500 --0.5 (2), 3250(5), 3110 (6), 3095(7),
3100 --0.0 (8), \\
& [Fe/H]=--0.5$\pm$0.3 & 3250 (9), 3110 (14), 3210 (15), 3130 (16), 3140 (17)\\
      GJ~406  & 3000$\pm$100 K,& 2800(5), 2580(6), 2670(7),
2600 --0.0 (8), 2850 (9), 2800 (10), 2500 (11), \\
& [Fe/H]=--0.5$\pm$0.5 & 2800 (13), 3000 (14), 2800 (15), 2565 (16), 2600 (17)\\      BS~8621 & $2800$ K & 3375--3475 (18) \\
\label{literature}
\end{tabular}
\end{minipage}
\end{table*}

\subsection{BS~8621}

BS~8621 is a giant star with a well determined spectral type of
M4III. Unlike the cool dwarfs previously considered it is 
likely that the oxygen (and 
carbon) abundances in its
atmosphere may be altered due to nuclear processing.
Nonetheless there are no indications in the literature
that this well studied object has a composition different
from solar so we start with the simplifying assumption
that oxygen and
carbon abundances in the atmosphere of BS~8621 are 
are solar: log
$N$(O)=--3.14 and log $N$(C)=--3.48 (Grevesse \& Anders 1989).

At a given temperature we expect that water bands should
weaken towards lower gravities as pressure broadening
lessens in importance. 
Although BS~8621 has a lower gravity and 
an early M spectral type, its spectrum has similar spectral features 
although a somewhat different spectral shape. As for the M dwarfs
its spectrum is clearly dominated by water vapour. However, the
spectrum is markedly different on the left-hand side from the
right-hand side. The longer wavelength side is well fit by 
a 2800K log $g$ = 3.0 model, see Fig.~\ref{f8621}. Though at 
wavelengths shorter than
about 2.7 $\mu$m, water vapour absorption does not completely dominate
the spectrum. It appears that transitions of CO $\Delta\nu$=2 
maybe important.
In order to explain the $<$ 2.7 $\mu$m spectrum with CO features it is 
necessary to considerably $increase$ the carbon abundance. This allows
the CO features to be more clearly seen against the background of 
water vapour features. 
In Fig.~\ref{ff8621} a synthetic spectrum with a carbon abundance
of log (C)~=~3.13 considerably improves the fit relative to
a synthetic spectrum with a solar abundance pattern (log (C)~=~--3.48,
log $N$(O)~=~--3.12). From the literature BS~8621 appears to be accepted
as a close to solar metallicity standard M4III though  
Lazaro et al. (1991) find that BS~8621 has a reduced carbon 
abundance [C/H]=--0.5. Lazaro et al.'s result thus supports 
the expectation that the C/O ratio in giants such as BS~8621 
may be considerably modified by the CN-cyle.
Such a drastic change in carbon
abundance appears to have a relatively small effect on the 
synthetic spectra longward of 2.7 $\mu$m (Fig.~\ref{f8621}) 
where CO bands are considerably weaker. In order
to properly investigate this we need to use structural models
computed for a variety of C/O ratios. Such models are not 
currently available to us and anyways are beyond the scope of 
the current paper where we wish to investigate the regime 
where water vapour is the dominant opacity, e.g. Fig.~\ref{3155a-m}.

\subsection{Discussion \label{discussion}}
We have investigated a variety of model fits for a range of M stars. 
These are particularly
promising for our cooler targets. In Table \ref{literature} we present
best fit parameters for our sample though it is important to note
that our study of the sensitivity of the spectra to the various 
model parameters is arguably more lasting than our determination 
of $T_{\rm eff}$, log $g$, and [Fe/H] values. We could have picked
out these values much more quickly by eye though we preferred  
do this systematically and thus be able to   
quantify the sensitivity of the spectra to different model parameters. 

In general the temperatures that 
we find are relatively high in comparison with the literature values
which were obtained with a variety of different methods.
An immediate concern is that our temperature for GJ~699 is higher 
than the 3100 K 
obtained for the benchmark eclipsing binary system CM~Dra composed of two
M4.5 stars (Viti et al. 1997). The bolometric
luminosity for GJ~699 is lower by around 20\% so it would be expected to have 
a lower temperature than CM~Dra. However, Viti et al. (2001)
have recently demonstrated that CM~Dra has a lower metallicity than
GJ~699 by around 0.5 dex. Based on Chabrier \& Baraffe (1997)
for a given luminosity this lower metallicity will cause a smaller 
radius by around 3\% or more which may compensate for the luminosity
difference allowing our temperature for GJ~699 to be just acceptable. 
One possibility is that the 
effective temperature measured by spectra from 2.5 --- 3.0 $\mu$m 
is not a good reflection of effective temperature of the spectral
energy distribution. Since the opacity of water vapour dominates
the infrared opacity budget and is particularly high in this region
we would expect to see little contribution at $\tau \sim$ 1 in this
region from deeper hotter layers thus we would actually expect to 
see cooler effective temperatures than expected but not hotter.
A more plausible explanation for the high temperatures we find 
is the non-physical
line splittings of the PS line list.
These are likely to over predict the strength of water vapour 
transitions for a given temperature. This will lead to higher 
temperatures being fit to a given observed spectrum.

We found very good fits for GJ~699 and GJ~406, however,
our fits for GJ~191 and BS~8621 are not so impressive.
We consider that the relatively poorer fit for GJ~191
arises because of its higher temperature though we attribute
our fitting problems with BS~8621 to the relative importance
of CO in its atmosphere.
For the case of GJ~191, our fit parameters are within the
wide spread of parameters found by other authors. In contrast
to the relatively high temperatures found for dwarfs our 
formal solution for the giant finds a significantly low 
temperature. We observe water vapour bands that are considerably 
stronger than are expected for its low gravity. Given that the
temperatures for M4III giants are determined
using interferometry results our 600 K lower temperature
are notable! 
However, our fit to BS~8621 does not account
properly for its apparent CO absorption bands.  The evidence from
Lazaro et al. (1991) is that BS~8621 is carbon deficient.  
A lower value of log $N$(C) is likely to dramatically alter
the chemical equilibrium, and as a result, CO and H$_2$O opacities
and the temperature structure of the model atmosphere. 
In particular the CO opacity will decrease with a corresponding increase
in H$_2$O opacity.  This increase in H$_2$O opacity is likely to 
be important accross a wide range of effective temperatures.
In the case of BS~8621 this increase in H$_2$O opacity will lead to an
increase in the relative strength of water bands for a given temperature
and thus cause us to fit BS~8621 to a higher temperature.
Furthermore, as discussed above, we expect this spectral region 
to yield relatively low temperatures. This idea will soon be 
tested as van Belle et al. (2001) are 
investigating the different effective temperatures found
when using effective radius measurements made at different
wavelengths from V to K. Another part of the solution to resolve 
the temperature discrepancy is likely to be provided by
the presence of a warm molecular sphere residing above the photosphere,
the so-called MOLsphere (Tsuji 2000a \& 2000b). However, given that
we do see CO bands from 2.5--2.7 $\mu$m it is not possible to 
properly test this scenario until spectral analysis has been
done with appropriate structural models and a variety of C/O ratios.

\section{Conclusions}

Observations of water vapour show a good match with previous
ground-based observations and indicate that the 
PS line list predicts the positions and
intensities of water vapour well enough to use for the
determination of effective temperatures. The PS line list is 
a substantial improvement on the MT line list which 
has been widely used for the generation of synthetic spectra 
of M dwarfs. The SCAN line list produced
results substantially different from
the observations and other models. The effective temperatures
we determined are reasonably consistent with other methods 
of temperature measurement.  Although the complex
spectral energy distributions of cool dwarfs caused by water
vapour have traditionally hampered reliable temperature
determinations, it now seems feasible that observations of water
rich regions coupled with the next generation of water vapour line
lists might become the method of choice for the temperature
determination of cool dwarfs.  Our work indicates that most sensitive
best fits will be obtained when analysing high resolution data.

However,
a further conclusion of this work, in agreement with that of 
Allard et al. (2000), is that the
presently available water linelists are still not good enough to generate
accurate spectroscopic models of cool stars. These linelists have however
been instrumental in improving the interpretation of hot water, e.g.
in sunspots (Polyansky et al. 1997b); as a result there are now significantly
more experimental data available on water (Tennyson et al. 2001). These
data are being used to greatly improve the effective potential energy
surfaces for water which, when combined with the improved dipole
surface due to Schwenke \& Partridge (2000), should make an excellent
starting point for generating a new linelist. Experience, such as the
tests performed in this paper, show that
all aspects of such calculations need to be of very high quality if a
satisfactory water opacity is to be obtained. 

\section*{Acknowledgments}

We thank Peter Hauschildt for synthetic spectra, David Schwenke
for water vapour data, David Goorvitch for CO data and
Iain Steele for organising the initial programme files
in Nordvijk, ESA.
We are also grateful to Kieron Leech and Alberto Salama for their
assistance with data reduction. 
Takashi Tsuji and the anonymous referre are warmly thanked
for their critical readings of this paper.  
ISO is an ESA project with instruments funded by ESA Member States
(especially the PI countries:
France, Germany, the Netherlands and the United Kingdom) and with
the participation of ISAS and NASA. The ISAP is a joint development by the LWS
and SWS Instrument Teams and Data Centers. Contributing institutes are CESR,
IAS, IPAC, MPE, RAL and SRON.
YPs studies are partially supported by a Small Research
Grant from American Astronomical Society.

\bsp

\label{lastpage}


\begin{thebibliography}{99}

\bibitem[\protect\astroncite{Allard}{2000}]{SPU_all73}
Allard F., Hauschildt P., 1999, ApJ, 540, 1005

\bibitem[\protect\astroncite{Allard}{2000}]{SPU_all73}
Allard F., Hauschildt P., Schwenke D.W., 2000, ApJ, 540, 1005

\bibitem[\protect\astroncite{Allen}{2000}]{Allen73}
Allen C.W., 1973, Astrophysical quantities, The Athlone Press,
University of London.

\bibitem[\protect\astroncite{Allard}{2000}]{SPU_all73}
Anders E., Grevesse N., 1989, Geochimica et Cosmochimica Acta,
53, 197   

\bibitem[\protect\astroncite{Allard}{2000}]{SPU_all73}
Basri G., Mohanty S., Allard F., Hauschildt P.H., Delfosse
X., Martin E.L., Forveille T., Goldman B., 2000, ApJ, 538, 363

\bibitem[\protect\astroncite{Allard}{2000}]{SPU_all73}
Berriman G., Reid N., Leggett S.K., 1992,  ApJ, 392, 31

\bibitem[\protect\astroncite{Allard}{2000}]{SPU_all73}
Bessell M.S., 1991,  AJ, 101, 662

\bibitem[\protect\astroncite{Allard}{2000}]{SPU_all73}
Brett J.M., 1995, A\&A, 295, 736

\bibitem[\protect\astroncite{Allard}{2000}]{SPU_all73}
Cayrel de Strobel G., Bentolila C.,
Hauck B., Duquennoy A., 1985, A\&AS, 59, 145

\bibitem[\protect\astroncite{Allard}{2000}]{SPU_all73}
Chabrier G. \& Baraffe I., 1997, A\&A, 327, 1039

\bibitem[\protect\astroncite{Goorvitch}{1994}]{Goorv94}
Goorvitch, D., 1994, ApJS, 95, 535

\bibitem[\protect\astroncite{Allard}{2000}]{SPU_all73}
de Graauw T. et al., 1996, SWS instrument manual, ESA
special publication

\bibitem[\protect\astroncite{Jones}{1996}]{SPU_vit97}
Gizis J., 1997, AJ, 113, 806

\bibitem[\protect\astroncite{Gurtovenko}{1996}]{SPU_vit97}
Gurtovenko E.A., Kostyk R.I., 1989, Fraunhofer spectrum and system of solar
oscillator strengths, Kiev, Izdatel'stvo Naukova dumka, 200


\bibitem[\protect\astroncite{Allard}{2000}]{SPU_all73}
Hauschildt P.H., Allard F., Baron E.,  1999, ApJ, 512, 377     
http://dilbert.physast.uga/edu/$\sim$yeti/mdwarfs.html

\bibitem[\protect\astroncite{Jones}{1996}]{SPU_vit97}
Heras, A.M., Shipman, R.F., Price, S.D., de Graauw, Th.,
Waters, L.B.F.M., Walker, H.J., de Muizon, M. Jourdain,
Kessler, M.F., Prusti, T, 1997, Astrophysics and Space Science,
255, 251

\bibitem[\protect\astroncite{Jones}{1996}]{SPU_vit97}
Jones H.R.A., Longmore A.J., Jameson R.F., Mountain C.M., 1994,
MNRAS, 267, 413

\bibitem[\protect\astroncite{Jones}{1995}]{SPU_all73}
Jones H.R.A., Longmore A.J., Hauschildt P., Allard F.A., Miller S.,
Tennyson J., 1995, MNRAS, 277, 767

\bibitem[\protect\astroncite{Jones}{1996}]{SPU_vit97}
Jones H.R.A., Longmore A.J., Hauschildt P., Allard F.A., 1996,
MNRAS, 280, 77

\bibitem[\protect\astroncite{Jones}{1996}]{SPU_vit97}
Jones H.R.A., Viti S., Miller S., Tennyson J., Hauschildt P., 1996,
9th Cambridge workshop on 
cool stars, stellar systems and the Sun, ASP, 109, 717, ed. Pallavicini, R.,
Dupree, A.K., Astronomical Society of the Pacific, San Francisco

\bibitem[\protect\astroncite{Jones}{1996}]{SPU_vit97}
Jorgensen U.G., Jensen P., Sorensen G.O., Aringer B., 2001, A\&A, 372, 249

\bibitem[\protect\astroncite{Allard}{2000}]{SPU_all73}
Kirkpatrick J.D., Kelly D.M., Rieke G.H., Liebert J., 
Allard F., Wehrse R., 1993, ApJ, 402, 643

\bibitem[\protect\astroncite{Allard}{2000}]{SPU_all73}
Krawchuk C.A.P., Dawson P.C.,
De Robertis M.M.,  2000, AJ, 119, 1956

\bibitem[]{_VALD_} 
Kupka, F., Piskunov, N., Ryabchikova, T. A., Stempels, H. C., 
Weiss, W. W., A\&AS, 1999, 138, 119

\bibitem[\protect\astroncite{Allard}{2000}]{SPU_all73}
Lazaro C., Lynas-Gray A.E., Clegg R.E.S., Mountain C.M.,
Zadrozny A., 1991, MNRAS, 249, 62

\bibitem[\protect\astroncite{Allard}{2000}]{SPU_all73}
Leggett S.K., 1992, ApJS, 82, 351

\bibitem[\protect\astroncite{Allard}{2000}]{SPU_all73}
Leggett S.K., Allard F., Dahn C., Hauschildt P.H.,
Kerr T.H., Rayner J., 2000, ApJ, 535, 965

\bibitem[\protect\astroncite{Jones}{1996}]{SPU_vit97}
Lutz D., Feuchtgruber H., Morfill J., 2000,
MPE-ISO-99-1

\bibitem[\protect\astroncite{Allard}{2000}]{SPU_all73}
Lynas-Gray A.E., Miller S., Tennyson J., 1995,
J. Mol. Spectrosc, 169, 458

\bibitem[\protect\astroncite{Jones}{1996}]{SPU_vit97}
Miller S., Tennyson J, Jones H.R.A., Longmore A.J.,
1994, in Proc. IAU Colloq. 146, Jorgensen U.G., Thejl Pl, eds, Springer-Verlag, Berlin, p. 296

\bibitem[\protect\astroncite{Jones}{1996}]{SPU_vit97}
Mould J.R., 1976, ApJ, 210, 402

\bibitem[\protect\astroncite{Allard}{1999}]{SPU_all73}
Partridge H., Schwenke D.W., 1997, J.Chem. Phys., 106, 4618

\bibitem[\protect\astroncite{YP}{2000}]{YP_00}
Pavlenko Y., 2000, Ast. Rep., 44, 219

\bibitem[\protect\astroncite{YP}{2001}]{P2001}
Pavlenko, Y. 2001, Ast. Rep., 45, 144 

\bibitem[\protect\astroncite{Jones}{1996}]{SPU_vit97}
Pettersen B.R., 1980, A\&A, 82, 53

\bibitem[\protect\astroncite{Jones}{1996}]{SPU_vit97}
Polansky O.L., Tennyson J., Viti S., Bernath P.F., 
Wallace L., 1997a, ApJ, 489, L205

\bibitem[\protect\astroncite{Jones}{1996}]{SPU_vit97}
Polansky O.L., Zobov N.F., Tennyson J., Viti S., 
Bernath P.F., Wallace L., 1997b, Journal 
Molecular Spectroscopy, 186, 422

\bibitem[\protect\astroncite{Jones}{1996}]{SPU_vit97}
Reid I.N., Gilmore G., 1984, MNRAS, 206, 19

\bibitem[\protect\astroncite{Jones}{1996}]{SPU_vit97}
Rothman L.S., Rinsland C.P., Goldman A., Massie S.T., Edwards D.P., Flaud J.-M., Perrin A., Camy-Peyret C., Dana B., Mandin J.-Y., Schroeder J., McCann A., Gamache R.R., Wattson R.B., Yoshino K., Chance K.V., Jucks K.W., Brown L.R., Nemtchinov V, Varanasi P., 1998, JQRST, 60, 1   

\bibitem[\protect\astroncite{Jones}{1996}]{SPU_vit97}
Salama A. et al. 2000, ESA SP-419 

\bibitem[\protect\astroncite{Jones}{1996}]{SPU_vit97}
Schryber J.H, Miller S., Tennyson J., 1994, JSQRT, 53, 373

\bibitem[\protect\astroncite{Jones}{1996}]{SPU_vit97}
Schwenke D.W., Partridge H., 2000, Journal of Chemical Physics, 
113, 6592)

\bibitem[\protect\astroncite{Jones}{1996}]{SPU_vit97}
Tennyson J., Zobov N.F., Williamson R., Polyansky O.L., 
Bernath P.F., 2001,
J. Phys. Chem. Ref. Data, in press

\bibitem[\protect\astroncite{Jones}{1996}]{SPU_vit97}
Tinney C.G., Mould J.R., Reid I.N., 1993, AJ, 105, 1045

\bibitem[\protect\astroncite{Allard}{1999}]{SPU_all73}
Tsuji T., Ohnaka K., Aoki W., 1996, A\&A, 305, L1

\bibitem[\protect\astroncite{Allard}{1999}]{SPU_all73}
Tsuji T., 2000, ApJ, 538, 801

\bibitem[\protect\astroncite{Allard}{1999}]{SPU_all73}
Tsuji T., 2000, ApJ, 540, 99

\bibitem[\protect\astroncite{Unsold}{1999}]{Unsold55}
Unsold, A. Physics der Sternatmospharen, 1955, Springer

\bibitem[\protect\astroncite{Jones}{1996}]{SPU_vit97}
Valentijn E.A., Feuchtgruber H., Kester D.J.M., 1996,
A\&A, 315, L60

\bibitem[\protect\astroncite{Jones}{1996}]{SPU_vit97}
van Belle G.T., Thompson R.R., PTI Collaboration, 2000, AAS,
197, 4502

\bibitem[\protect\astroncite{Allard}{2000}]{SPU_all73}
van den Ancker M.E., Voors R., Leech K., 1997, SWS internal
report, 1.7.1997

\bibitem[\protect\astroncite{Jones}{1996}]{SPU_vit97}
Veeder G., 1974, AJ, 79, 1056

\bibitem[\protect\astroncite{Viti}{1997}]{SPU_vit97}
Viti S., 1997, PhD thesis, University of London

\bibitem[\protect\astroncite{Viti}{1997}]{SPU_vit97}
Viti S., Tennyson J., Polyansky O.L., 1997, MNRAS, 287, 79

\bibitem[\protect\astroncite{Viti}{1997}]{SPU_vit97}
Viti S., Jones H.R.A., et al., 1997, MNRAS, 287, 79

\bibitem{b25}
Viti S., Jones H.R.A, 1999, A\&A, 351, 1028

\bibitem[\protect\astroncite{Jones}{1996}]{SPU_vit97}
Viti S., Jones H.R.A., Maxted P.F.L., Tennyson J., 2001,
MNRAS, submitted

\bibitem[\protect\astroncite{Jones}{1996}]{SPU_vit97}
Zuckerman B., Dyck H.M., 1986, ApJ, 311, 345

\end{thebibliography}
\end{document}